\documentclass[12pt]{article}

\def\ergscm2 {erg\,s$^{-1}$cm$^{-2}$}

\usepackage{graphicx}
\usepackage{url}
\usepackage{color}

\usepackage{times}

\title{A highly resistive layer within the crust of X-ray pulsars limits their spin periods}

\author
{Jos\'e A. Pons$^{1}$, Daniele Vigan\`o$^{1}$ and Nanda Rea$^{2}$\\
\normalsize{$^{1}$ Departament de F\'{\i}sica Aplicada, Universitat d'Alacant,}\\ 
\normalsize{Ap. Correus 99, 03080, Alacant, Spain}\\
\normalsize{$^{2}$ Institut de Ci\`encies de l'Espai (CSIC-IEEC),}\\
\normalsize{Campus UAB, Facultat de Ci\`encies, Torre C5-parell, E-08193 Barcelona, Spain}\\
}

\date{}

\begin{document}


\baselineskip24pt


\maketitle

\begin{abstract}

The lack of X-ray pulsars with spin periods $>12$ s raises the question of where 
the population of evolved high magnetic field neutron stars has gone. 
Unlike canonical radio-pulsars, X-ray pulsars are not subject to physical limits to the emission mechanism nor 
observational biases against the detection of sources with longer periods. 
Here we show that a highly resistive layer in the innermost part of the crust of neutron stars naturally limits the spin period
to a maximum value of about 10-20 s. This highly resistive layer is expected if
the inner crust is amorphous and heterogeneous in nuclear charge, possibly due to the existence of a nuclear pasta phase.
Our findings suggest that the maximum period of isolated X-ray pulsars can be the first observational
evidence of an amorphous inner crust, which properties can be further constrained by future X-ray timing missions
combined with more detailed models.

\end{abstract}

\newpage

It has long been hoped that stringent constraints could be placed 
on the equation of state of dense matter from astrophysical measurements of neutron stars, 
the compact remnants of the explosion of massive stars. 
Up to the present time, however, despite important breakthrough discoveries
\cite{dem10,vk11,sht11} we still fall short (see e.g. \cite{ste10} for a recent review).
The majority of neutron stars  are observed as radio-pulsars  with magnetic
fields in the $\sim 10^{11}-10^{13}$~G range, that spin down due to magneto-dipole radiation losses, 
while converting a fraction of their rotational  energy into electromagnetic radiation.  
The traditional wisdom that, when the star is not accreting matter from a companion star in a binary system, 
its main energy source comes from rotational energy, 
was shattered with the discovery of the so-called {\it magnetars}, a class that includes 
anomalous X-ray pulsars and soft gamma-ray repeaters (SGRs) \cite{mer08}.
At present, there are over twenty of these (mostly young) neutron stars characterized by high X-ray
quiescent luminosities (generally larger than their entire reservoir
of rotational energy), short X-ray bursts \cite{gavr02}, sometimes displaying giant flares
\cite{pal05}, and/or transient pulsed radio emission \cite{cam06}.

In the most successful model {\cite{dt92},  
magnetars are believed to be endowed with large magnetic
fields, $B\sim 10^{14}-10^{15}$~G, which explains their large periods ($2-12$ s), compared to
those of normal isolated radio-pulsars (mostly in the range $0.1-1$ s). Relatively high fields are also 
observed in a class of nearby
thermally emitting neutron stars usually known as {\it the magnificent seven}.  These are objects with 
magnetic fields of  $\sim 10^{13}$~G, intermediate between
magnetars and radio-pulsars, but their periods also cluster in the same range as for magnetars.
Although there are hints pointing to an evolutionary link between magnetars, nearby isolated X-ray pulsars, 
and some high-B radio-pulsars, a complete "grand unification theory" is still lacking \cite{kaspi}. 
Interestingly, some sources (e.g. 1E\,1841--045, SGR\,0526--66, SGR\,1806--20 among others) are known to be young and they already have periods from 7 to 12\,s . With their 
current estimated dipolar fields, they should easily reach periods of 20 or 30\,s in a few thousand more years. 
This seems in contradiction with a steady pulsar spin down rate. 
Where is the population of evolved high magnetic field neutron stars with long periods? 
Why none of the middle age magnetars, or the older X-ray pulsars have longer periods?

For rotation-powered radio-pulsars, the strong dependence of the radio luminosity and the beaming angle with the rotation period, and the selection effect of several radio surveys for long spin periods, result in the lack of observed radio pulsars with periods longer than a few seconds. For X-ray pulsars, however, there is no reason to expect any selection effect. 
We plot in Figure \ref{fig1} the spin period distribution of isolated X-ray pulsars and X-ray binary pulsars, showing that there are no observational limitations to the detection of slow periods in X-ray binaries. When other torques are present (accretion), X-ray pulsars
with rotation periods of hundreds or even thousand of seconds are clearly observed.
The fact that no X-ray emitting isolated neutron star has been discovered so far with a period $>12$s must
therefore be a consequence of a real physical limit  and not simply a statistical fluctuation \cite{psa02}.
The easiest and long-standing answer is that the magnetic field decays as the neutron star gets older \cite{col00}
in such a way that its spin-down rate becomes too slow to lead to longer rotation periods during the time it is still bright enough to
be visible as an X-ray pulsar. In this scenario, 
{\it low-field magnetars} \cite{rea} are simply old magnetars whose external dipolar magnetic field has decayed to normal values
\cite{tur11,rea12}. 
However, no detailed quantitative predictions supported by realistic simulations
have been able to reproduce the observational limits. This is the purpose of this work.

The long-term life in the interior of a neutron star is very dynamic.  
As the star cools down, the internal magnetic field is subject to a continuous
evolution through the processes of Ohmic dissipation, ambipolar diffusion, and Hall drift \cite{gol92,cum04}.
Soon after birth (from hours to days, at most) protons in the liquid core undergo a transition to a 
superconducting phase and a solid crust is formed. 
The neutron star crust (see \cite{cha08} for a comprehensive review) 
is for the most part an elastic solid, comprising a Coulomb lattice of normal spherical nuclei. 
It is only about 1\,km thick (10\% of the star radius) and contains 1\% of its
mass, however it is expected to play a key role in various observed astrophysical phenomena 
(pulsar glitches \cite{and75}, quasi-periodic oscillations in SGRs \cite{stro06}, 
thermal relaxation in soft X-ray transients \cite{bro09}, etc.)
Among all the ingredients that determine the magneto-rotational evolution of
a neutron star, one key issue is the magnetic diffusivity in the region where currents are placed.
If currents supporting the magnetic field of neutron stars are predominantly in the core,
the extremely high conductivity would result in little field dissipation, thus keeping a nearly constant magnetic 
field during the first Myr of a pulsar life.
On the contrary, a magnetic field supported by currents in the crust is expected to appreciably decay in $\approx 1$ Myr,  
because of the combined action of Ohmic dissipation and the Hall drift \cite{pons07,vig12}.
Therefore, the crust is actually the most important region to determine the magnetic field dissipation 
timescale.

The inner crust is where the {\it nuclear pasta phase}, a novel state of matter having nucleons arranged 
in a variety of complex shapes, is expected to appear.
In the innermost layers near the crust/core boundary, because of the large effect of the Coulomb lattice energy, cylindrical 
and planar geometries can occur, both as nuclei and as bubbles \cite{rav83}. These phases
are collectively named nuclear pasta (by analogy to the shape of spaghetti, maccaroni and lasagna). 
More sophisticated molecular dynamics simulations \cite{hor05,hor08} have shown that it
may be unrealistic to predict the exact sizes and shapes of the pasta
clusters, and that the actual  shape of the pasta phase can be very amorphous, with a very irregular 
distribution of charge. This  is expected to have a strong impact on the thermodynamic and transport properties \cite{wat00}, in particular on the electrical resistivity, because electron scattering off crystal impurities becomes the dominant process when the star has cooled down enough (typically, $\sim 10^5$ yr).
The relevant nuclear parameters that determine the thickness of the crust, 
the range of densities at which pasta might appear, or the concentration of impurity charges are the 
symmetry energy and its density dependence close to nuclear saturation density. 
This important parameters also determine global properties such as the radius and moment of inertia, and have been proposed to have a potential observational effect in the crust oscillation 
frequencies \cite{sot11,new11}. 
Even assuming spherical nuclei, the inner crust has also been proposed to be amorphous and 
heterogeneous in nuclear charge \cite{jon04,jon04b},
with electrical and thermal conductivities much smaller than for a homogeneous body-centered cubic lattice.
A particularly important issue is the problem of shell effects associated with
unbound neutrons scattered by nuclear inhomogeneities \cite{mag02}, which seems to result in the coexistence of
several phases different from a bcc lattice.
However, we note that there are also open issues about phase separation and recent molecular dynamics simulations seem
to favor a regular bcc lattice, even in accreting systems, when a large number of impurities are present \cite{hor09a,hor09b}.
Recent studies of diffusion (motion of defects) \cite{hor11} suggest that the
the crust of neutron stars will be crystalline and not amorphous, at least for the outer crust.

To investigate the observable effects of the different possibilities, we adopt the impurity parameter formalism as a first
simplified approximation to the complex calculation of the resistivity.
The impurity parameter, $Q_{imp}$, is a measure of the distribution of the nuclide charge numbers in the crust material,
and it tells how ``crystalline" ($Q_{imp}\ll 1$) or ``disordered"  ($Q_{imp}\gg 1$) is the neutron star crust.
A pasta region, by definition, is expected to have a large  $Q_{imp}$
since it is expected that clustering happens in a very irregular manner. 
We also note that the conductivity of an amorphous crust calculated with molecular dynamics simulations
is even lower, by an order of magnitude, than the estimates obtained using the impurity parameter
formalism (as shown in \cite{dal09} for the outer crust in the scenario of accreting neutron stars). 
The ``effective'' $Q_{imp}$ to be used in the impurity parameter formalism must then 
be high to reproduce the molecular dynamics results.
We will set $Q_{imp}=0.1$ in the outer crust, but let it vary in the inner crust, in order to 
phenomenologically explore the sensitivity of our results to this important parameter.
We consider four models (A, B, C, D) with a high $Q_{imp}$ only in the pasta region, $\rho > 10^{13}$ g/cm$^3$, 
but low $Q_{imp}$ in the rest of the inner crust, one model (E) with $Q_{imp}=0.1$ everywhere, and another model (J) corresponding to the extrapolation of the values 
calculated in \cite{jon04,jon04b} at a few densities. The profiles of $Q_{imp}$ as a function of density and the corresponding profiles of
electrical resistivity are shown in Fig. \ref{fig2}, and the summary of the properties of each model is given in Table~\ref{tab1}.

As initial model, we have chosen a neutron star born with an initial dipolar field of $B=3\times 10^{14}$ G (at the pole). 
We refer to section 2 in \cite{pons09}, section 4 of \cite{agui08}, and references therein, for all details about the magnetic field geometry, the equation of state employed and other microphysical inputs. 
The evolution is followed for $10^6$~years using the new 2D code developed
in \cite{vig12}, to which we refer for technical details.
In Fig.~\ref{fig3} we show the value of the magnetic field at the pole as a function of age for the different models.
During the first $\sim$ 30000 years, the field is dissipated by a factor $\sim 2$ for all the models, with slight differences due to the different neutron star masses and $\Delta R_{crust}$. 
In this regime, the neutron star crust is still warm and electron scattering off impurities is not the dominant process. 
Thereafter, as the star cools, the evolution strongly depends on the impurity parameter in the inner crust. 
For low values of $Q_{imp}$ (model E), the field almost stops to dissipate and remains high, with some oscillations due to the Hall
term in the induction equation \cite{pons07,vig12}. By contrast, a large value of $Q_{imp}$ (models A, B, C) results in the dissipation
of the magnetic field by one or even two orders of magnitude between 0.1 and 1 Myr. 
Models D and J, both with moderated values of $Q_{imp}$, show an intermediate dissipation rate.
These differences in the time evolution of the magnetic field have a clear observational effect that we discuss now. 

The spin--down behavior of a rotating neutron star is governed by the energy balance equation relating the loss of rotational energy to the energy loss rate $ \dot{E}$, given by magneto-dipole radiation, wind, gravitational radiation, or other mechanisms. The most accurate calculations of the spin-down luminosity of an oblique rotator \cite{Li12} can be well approximated by
\begin{equation}
\dot{E}= \frac{B_p^2R^6\Omega^4}{4 c^3} (1+\sin{\alpha}^2)\;,
\label{equ:mdr}
\end{equation}
where $R$ denotes the neutron star radius, $\Omega=2 \pi /P$ is the angular velocity,
$\alpha$ is the angle between the rotational and the magnetic axis, and $c$ is the speed of light. The energy balance
equation between radiation  and rotational energy losses gives
\begin{equation}
\label{spindown}
I\Omega \dot{\Omega} =  \frac{B_p^2R^6\Omega^4}{4 c^3} (1+\sin{\alpha}^2)
\end{equation}
where $I$ is the effective moment of inertia of the star. 

Numerically integrating equation (\ref{spindown}) with $B_p(t)$ obtained  from our simulations, we can obtain
evolutionary tracks in the $P-\dot{P}$ diagram, which allows a close comparison
to observed timing properties of X-ray pulsars. This is shown in Fig.~\ref{fig4}, where we compare the theoretical 
trajectories of a neutron star up to an age of 10\,Myr to the sample of isolated X-ray pulsars with the largest rotation periods.
The most important qualitative difference is that, for models with high $Q_{imp}$, the evolution tracks become
vertical after $\sim 10^5$ yr, indicating that the period has reached an asymptotic value while its derivative steadily decreases. 
The particular upper limit of $P$ depends on the neutron star mass, the initial field and the value of $Q_{imp}$, 
but this gives a natural explanation to the observed upper limit to the rotation period of isolated X-ray pulsars, and the observed
distribution with objects of different classes  clustering in the range $P=2-12$ seconds while $\dot{P}$ varies over six orders of magnitude.
Models J and D (moderate impurity parameter) follow similar trajectories to models A-B-C, but reaching slightly longer periods at late times.
Conversely, in model E (low impurity in the pasta region) the period keeps increasing due to the slower dissipation of the magnetic field, which in principle predicts that pulsars of longer periods (20-100 s) should be visible. 
In models D and E, the slow release of magnetic energy through Joule heating keeps the neutron star bright and visible much longer than 
for the rest of models. The luminosity of models D and E, at an age of a few Myrs, is $\approx 10^{32}$ erg/s, high enough
for sources a few kpc away to be detectable with present X-ray instruments. 

Other possible torque mechanisms, that would enter as extra-terms in equation (\ref{spindown}), such as stellar wind or accretion from a
fallback disk, could act only in the early stages of a neutron star life and may contribute to explain the observed large values of $\dot{P}$ in some objects. Furthermore, the effective moment of inertia  $I$, may also vary with time as the superfluid part of the core grows during the
first hundreds of years. A simple phenomenological model for the rotational evolution of the normal and superfluid components and its observational implications have been discussed in \cite{ho12}. Note, however, that the main conclusion of our calculations is not affected:
if there is a highly disordered inner crust, either due to the existence of the pasta phase or because the whole inner crust is amorphous, evolutionary tracks in the  $P-\dot{P}$ diagram will bend down after $10^5$ yrs regardless of the particular model. As a consequence, there will be a maximum spin period for isolated X-ray pulsars. We also note that, under these conditions, the resistivity is 
dominated by electron-impurity scattering, which is almost independent of temperature (see right panel of Fig.~\ref{fig2}). Therefore, 
the effect of varying the superfluid gap, or other microphysical parameters that modify the local temperature at
different ages, becomes irrelevant.

We conclude that there is a direct correlation between the maximum observed
spin period of isolated X-ray pulsars and the existence of a highly resistive layer in the inner crust of
neutron stars. This can be interpreted as evidence for the existence of the pasta phase near the crust/core interface or 
as evidence for a highly disordered whole inner crust. The current development of theoretical models and the existing data
do not allow to distinguish between this two options, but the imprint of the transport properties of the inner crust is
clearly visible in the timing properties of the neutron star population. This opens a new possibility of further constraining the transport
properties of the crust using the distributions of the spin period and period derivative.
In particular, as present and future space missions, such as LOFT (Large Observatory For X-ray Timing \cite{loft}),  
keep increasing the statistics of X-ray pulsars, and realistic theoretical models are used as input 
for neutron star population synthesis studies, we will be able to accurately
constrain the properties of the inner crust of neutron stars and, therefore, the equation of state
of dense matter.

%%%%%%%%%%%%%%%%%%%%%%%%%%%%%%%%%%%%%%%%%

\noindent {\bf Corresponding author:}\\
Correspondence to Jos\'e A. Pons

\noindent {\bf Acknowledgments}\\
This work has been supported by the grants AYA 2010-21097-C03-02,
AYA2012-39303, SGR2009-811, TW2010005 and iLINK 2011-0303.
NR is supported by a Ramon y Cajal Research Fellowship and DV by the Prometeo/2009/103 grant.

\noindent {\bf Author Contribution}\\
J.A.P. and D.V. contributed to developing the model, performed the calculations and wrote the manuscript. 
N.R. contributed to writing the manuscript and to select and check the observational data.

\newpage

%%%%%%%%%%%%%%%%%%%%%%
\begin{table}[ht!]
\begin{tabular}{l c c c c c}
\hline
\hline
Model & $M [M_\odot]$ & $I_{45}$ & $\Delta R_{crust}$ [km] & $\Delta R_{pasta}$ [km] & $Q_{max}$ \\
\hline
A	& 1.10	& 0.962 & 0.94 & 0.14 & 100   \\
B	& 1.40	& 1.327	& 0.70 & 0.10 & 100   \\
C	& 1.76	& 1.755 & 0.43 & 0.07 & 100   \\
D	& 1.40	& 1.327 & 0.70 & 0.10 &  10   \\
E	& 1.40	& 1.327 & 0.70 & 0.10 &   0.1 \\
J	& 1.40	& 1.327 & 0.70 & 0.0  &  23   \\ 
\hline
\hline
\end{tabular}
\caption{Summary of the properties of the neutron star models considered in this work: mass (in solar masses), moment of inertia (in units of $10^{45}$ g cm$^{-2}$), thickness of the crust and of the pasta phase, and maximum value of the impurity parameter.}
\label{tab1}
\end{table}
%%%%%%%%%%%%%%%%%%%%%%%

%%%%%%%%%%%%%%%%%%%%%
\begin{figure}
\includegraphics[width=16cm]{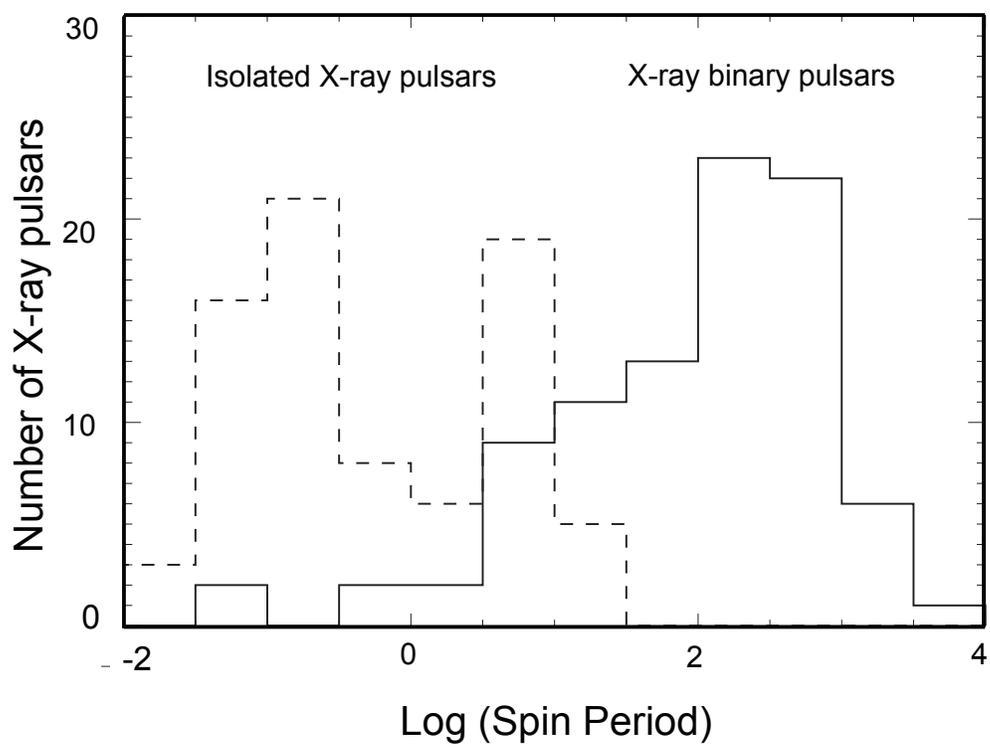}
\caption{Period distribution of isolated neutron stars (red) and neutron stars in binary systems (black).}
\label{fig1}
\end{figure}

%%%%%%%%%%%%%%%%%%%%%

%%%%%%%%%%%%%%%%%%%%%
\begin{figure*}
\includegraphics[width=16cm]{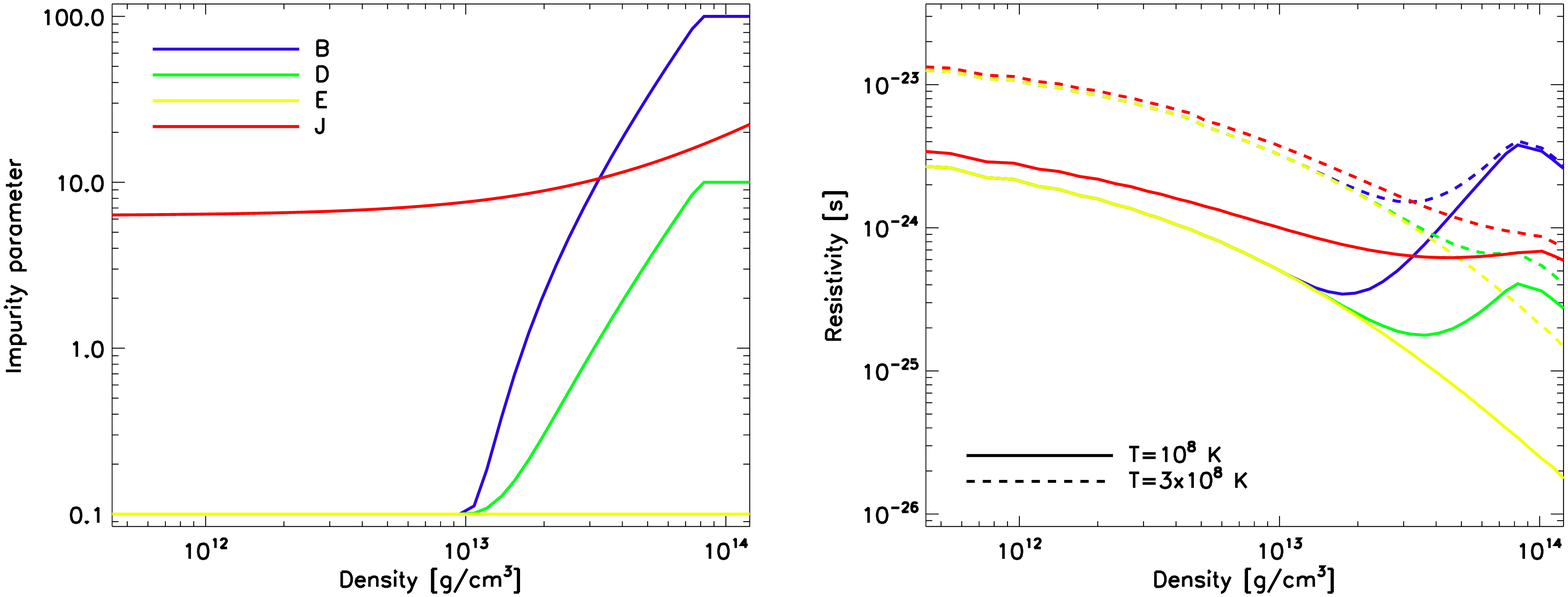}
\caption{Impurity parameter $Q_{imp}$ (left) and electrical resistivity (right) as a function of density for the four
models with $M=1.4 M_\odot$. We plot the resistivity for two different temperatures to show more clearly the regions where the
temperature-independent disorder resistivity dominates.}
\label{fig2}
\end{figure*}

%%%%%%%%%%%%%%%%%%%%%

%%%%%%%%%%%%%%%%%%%%%
\begin{figure}[ht!]
\includegraphics[width=16cm]{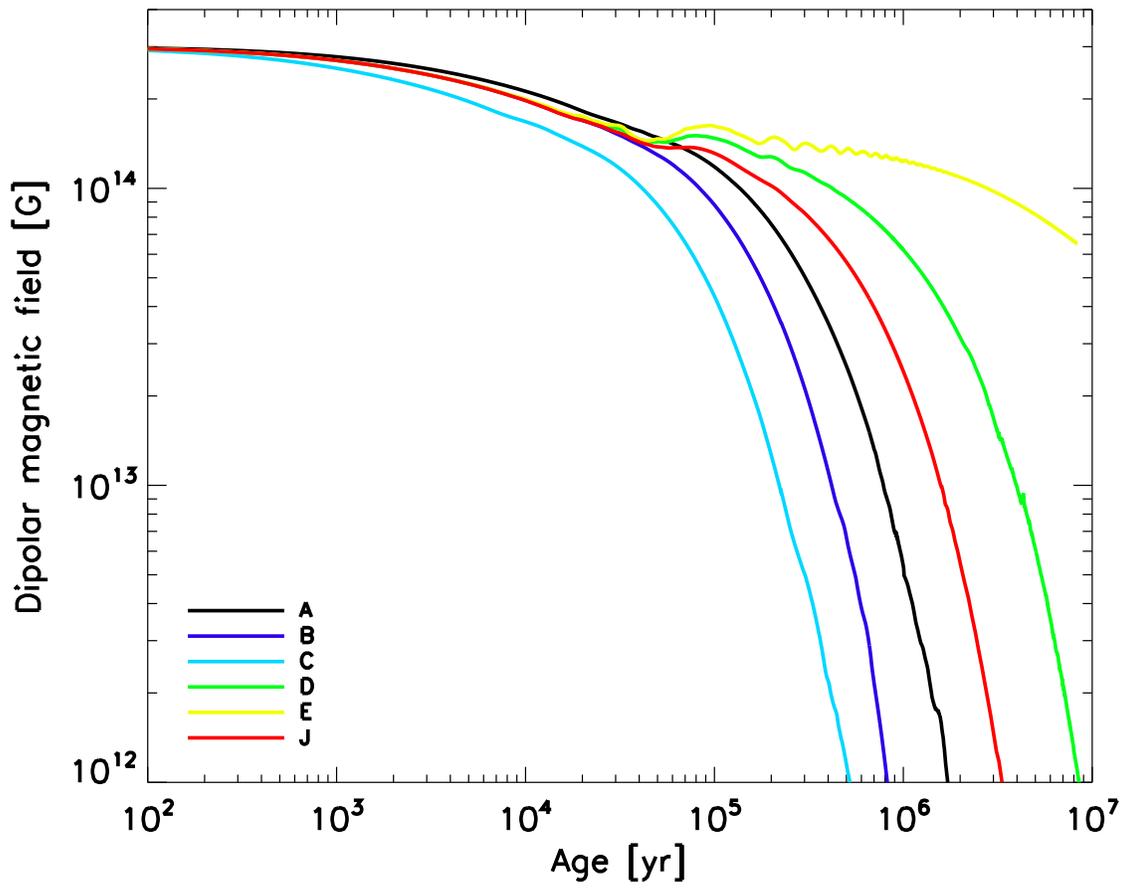}
\caption{Magnetic field strength at the pole ($B_{\rm p}$) as a function of the neutron star age for the models listed in Table~\ref{tab1}.}
\label{fig3}
\end{figure}
%%%%%%%%%%%%%%%%%%%%%
\begin{figure}[ht!]
\includegraphics[width=16cm]{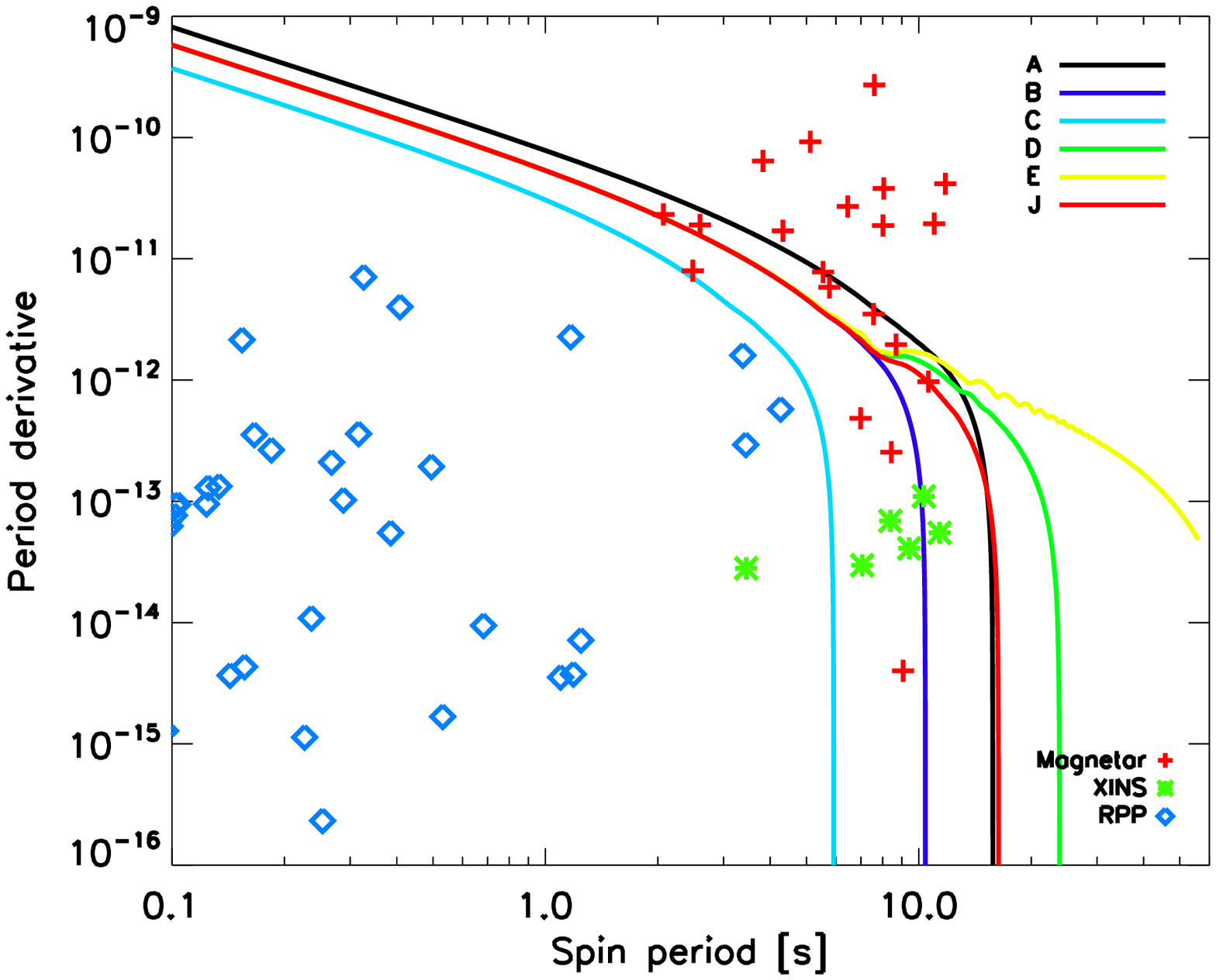}
\caption{$P-\dot{P}$ diagram for magnetars, X-ray Isolated Neutron Stars and Rotation Powered Pulsars with X-ray emission. 
We overplot evolutionary tracks for the six models listed in Table~\ref{tab1}.}
\label{fig4}
\end{figure}

%%%%%%%%%%%%%%%%%%%%%


\begin{thebibliography}{}

\bibitem{dem10}
{{Demorest}, P.~B., {Pennucci}, T., {Ransom}, S.~M., {Roberts}, M.~S.~E. \& {Hessels}, J.~W.~T.}
{A two-solar-mass neutron star measured using Shapiro delay.}
{\it Nature} {\bf 467}, 1081-1083 (2010).

\bibitem{vk11}
{{van Kerkwijk}, M.~H., {Breton}, R.~P. \& {Kulkarni}, S.~R.}
{Evidence for a Massive Neutron Star from a Radial-velocity Study of the Companion to the Black-widow Pulsar PSR B1957+20.}
{\it Astrophys. J} {\bf 728}, 95-102 (2011).

\bibitem{sht11}
{{Shternin}, P.~S. and {Yakovlev}, D.~G. and {Heinke}, C.~O. and {Ho}, W.~C.~G. \& {Patnaude}, D.~J.}
{Cooling neutron star in the Cassiopeia A supernova remnant: evidence for superfluidity in the core.}
{\it Mon. Not. R. Astron. Soc.} {\bf 412}, L108-L112 (2011).

\bibitem{ste10}
{{Steiner}, A.~W., {Lattimer}, J.~M. \& {Brown}, E.~F.}
{The Equation of State from Observed Masses and Radii of Neutron Stars.} 
{\it Astrophys. J} {\bf 722}, 33-54 (2010).	

\bibitem{mer08}	
{{Mereghetti}, S.}
{The strongest cosmic magnets: soft gamma-ray repeaters and anomalous X-ray pulsars.}
{\it The Astronomy and Astrophysics Review} {\bf 15}, 225-287 (2008).	

\bibitem{gavr02}
{{Gavriil}, F.~P., {Kaspi}, V.~M. \& {Woods}, P.M.}
{Magnetar-like X-ray bursts from an anomalous X-ray pulsar. }{\it Nature} {\bf 419}, 142-144 (2002).

\bibitem{pal05}{Palmer, D.M. et al.}
{A giant {$\gamma$}-ray flare from the magnetar SGR 1806 - 20.}
{\it Nature} {\bf 434}, 1107-1109 (2005).

\bibitem{cam06}{Camilo, F. et al.}
{Transient pulsed radio emission from a magnetar.}
{\it Nature} {\bf 442}, 892-895 (2006).

\bibitem{dt92} Duncan, R.C.  \&  Thompson, C.
Formation of very strongly magnetized neutron stars - Implications for gamma-ray bursts. 
{\it Astrophys. J} {\bf 392}, L9-L13 (1992).

\bibitem{kaspi} Kaspi, V.M. Grand unification of neutron stars
{\it Publ. Nat. Academy of Science}  {\bf 107}, 7147-7152 (2010).

\bibitem{rea} Rea, N. et al. A Low-Magnetic-Field Soft Gamma Repeater.
{\it Science} {\bf 330}, 944-946 (2010). 

\bibitem{tur11} Turolla, R., Zane, S., Pons, J.A., Esposito, P. \& rea, N.
Is SGR 0418+5729 Indeed a Waning Magnetar?
{\it Astrophys. J.} {\bf 740}, 105-111 (2011).

\bibitem{rea12} Rea, N. et al. 
A New Low Magnetic Field Magnetar: The 2011 Outburst of Swift J1822.3-1606
{\it Astrophys. J.} {\bf 754}, 27-40 (2012).

\bibitem{psa02}{{Psaltis}, D. \& {Miller}, M.~C.}
{Implications of the Narrow Period Distribution of Anomalous X-Ray Pulsars and Soft Gamma-Ray Repeaters}.
{\it Astrophys. J.} {\bf 578}, 325-329 (2002).

\bibitem{col00}
{{Colpi}, M., {Geppert}, U. \& {Page}, D.} 
{Period Clustering of the Anomalous X-Ray Pulsars and Magnetic Field Decay in Magnetars.}
{\it Astrophys. J} {\bf 529}, L29-L32 (2000).	

\bibitem{gol92} {{Goldreich}, P. \& {Reisenegger}, A.} {Magnetic field decay in isolated neutron stars.}
{\it Astrophys. J.} {\bf 395}, 250-258 (1992).
 
\bibitem{cum04}
{{Cumming}, A., {Arras}, P., \& {Zweibel}, E.}
{Magnetic Field Evolution in Neutron Star Crusts Due to the Hall Effect and Ohmic Decay.}
{\it Astrophys. J.} {\bf 609}, 999-1017 (2004).

\bibitem{cha08}
{{Chamel}, N. \& {Haensel}, P.}
{Physics of Neutron Star Crusts},
{\it Living Reviews in Relativity} {\bf 11} (2008) 10.

\bibitem{and75}
{{Anderson}, P.~W. \& {Itoh}, N.}
{Pulsar glitches and restlessness as a hard superfluidity phenomenon.}
{\it Nature} {\bf 256}, 25-27 (1975).

\bibitem{stro06}{{Strohmayer}, T.~E. \& {Watts}, A.~L.}
{The 2004 Hyperflare from SGR 1806-20: Further Evidence for Global Torsional Vibrations.}
{\it Astrophys. J} {\bf 653}, 593-601 (2006).

\bibitem{bro09} {{Brown}, E.~F. \& {Cumming}, A.}
{Mapping Crustal Heating with the Cooling Light Curves of Quasi-Persistent Transients.}
{\it Astrophys. J.} {\bf 698}, 1020-1032 (2009).

\bibitem{pons07} Pons, J.A. \& Geppert, U. 
{Magnetic field dissipation in neutron star crusts: from magnetars to isolated neutron stars.} 
{\it Astron. Astrophys} {\bf 470}, 303-315 (2007). 

\bibitem{vig12} Vigan\`o, D., Pons, J.A. \&  Miralles, J.A. {A new code for the Hall-driven magnetic evolution of neutron stars.}  
{\it Computer Phys. Comm.} {\bf 183}, 2042-2053 (2012).

\bibitem{rav83} 
{{Ravenhall}, D.~G., {Pethick}, C.~J. \& {Wilson}, J.~R.}
{Structure of Matter below Nuclear Saturation Density.}
{\it Phys. Rev. Lett.} {\bf 50}, 2066-2069 (1983).

\bibitem{hor05} 
{{Horowitz}, C.~J., {P{\'e}rez-Garc{\'{\i}}a}, M.~A., {Berry}, D.~K. \& {Piekarewicz}, J.}
{Dynamical response of the nuclear pasta in neutron star crusts.}
{\it Phys. Rev. C}  {\bf 72}, 035801 (2005).

\bibitem{hor08} 
{{Horowitz}, C.~J. \& {Berry}, D.~K.}
{Shear viscosity and thermal conductivity of nuclear pasta.}
{\it Phys. Rev. C}  {\bf 78}, 035806 (2008).

\bibitem{wat00} 
{Watanabe, G., Iida, K. \& Sato. K.}
{Thermodynamic properties of nuclear pasta in neutron star crusts.}
{\it Nuclear Physics A} {\bf 676}, 455-473 (2000)

\bibitem{sot11}{{Sotani}, H.}
{Constraints on pasta structure of neutron stars from oscillations in giant flares.}
{\it Mon. Not. R. Astron. Soc.} {\bf 417}, L70-L73 (2011).

\bibitem{new11} Gearheart, M., Newton, W.G., Hooker, J. \& Li, Bao-An. 
{Upper limits on the observational effects of nuclear pasta in neutron stars.}
{\it Mon. Not. R. Astron. Soc.}  {\bf 418}, 2343-2349 (2011).

\bibitem{jon04} Jones, P.~B.
{Disorder Resistivity of Solid Neutron-Star Matter}
{\it Phys. Rev. Lett.} {\bf 93}, 221101 (2004).

\bibitem{jon04b} Jones, P.~B.
{Heterogeneity of solid neutron-star matter: transport coefficients and neutrino emissivity.}
{\it  Mon. Not. R. Astron. Soc.} {\bf 351}, {956-966} (2004).

\bibitem{mag02} 
{Magierski, P. \& Heenen, P. H.}
{Structure of the inner crust of neutron stars: Crystal lattice or disordered phase?}
{\it Phys. Rev. C} {\bf 65}, 045804 (2002).

\bibitem{hor09a} 
{{Horowitz}, C.~J. \& {Berry}, D.~K.}
{Structure of accreted neutron star crust.}
{\it Phys. Rev. C}  {\bf 79}, 065803 (2009).

\bibitem{hor09b} 
{{Horowitz}, C.~J., Caballero, O.~L. \& {Berry}, D.~K.}
{Thermal conductivity and phase separation of the crust of accreting neutron stars.}
{\it Phys. Rev. E}  {\bf 79}, 026103 (2009).

\bibitem{hor11} 
{Hughto, J., Schneider, A. S., Horowitz, C. J., \& Berry, D. K.}
{Diffusion in Coulomb crystals.}
{\it Phys. Rev. E}  {\bf 84}, 016401 (2011).

\bibitem{dal09}
{Daligault, J. \& Gupta, S.}
{Electron-Ion Scattering in Dense Multi-Component Plasmas: Application to the Outer Crust of an Accreting Neutron Star.}
{\it Astrophys. J.}  {\bf 703}, 994-1011 (2009).

\bibitem{pons09}  Pons, J. A., Miralles, J.A., \& Geppert, U.  {Magneto-thermal evolution of neutron stars.}
{\it Astron. Astrophys} {\bf 496}, 207-216 (2009).

\bibitem{agui08} {Aguilera, D.N., Pons, J.A. \& Miralles , J.A.} 
{The Impact of Magnetic Field on the Thermal Evolution of Neutron Stars.}
{\it Astrophys. J.} {\bf 673}, L167-L170 (2008).

\bibitem{Li12} {{Li}, J., {Spitkovsky}, A. \& {Tchekhovskoy}, A.}
{Resistive Solutions for Pulsar Magnetospheres.} {\it Astrophys. J.} {\bf 746}, 60 (2012).

\bibitem{ho12} {Ho, W.C.G \& Andersson, N.} {Rotational evolution of young pulsars due to superfluid decoupling.}
{\it Nature Physics} {\bf 8}, 787-789 (2012).

\bibitem{loft} Feroci, M. et al. {The Large Observatory for X-ray Timing (LOFT).}
{\it Experimental Astronomy} {\bf 34}, {415-444} (2012).


\end{thebibliography}
\end{document}